\documentclass{astlb}

\usepackage{mathptmx}
\usepackage{txfonts}
\usepackage{pdflscape}
\usepackage[T1]{fontenc}
\usepackage{ae,aecompl}
\usepackage{graphicx}
\usepackage{changes}
\usepackage[hyperfootnotes=false]{hyperref}

\newcommand{\rx}{RX~J0535.0-6700}

\newcommand{\srg}{\textit{SRG}\,}

\def\lum{erg\,s$^{-1}$}
\def\flx{erg\,cm$^{-2}$\, s$^{-1}$}

\begin{document}
\journalinfo{2025}{51}{1}{0}[0]

   \title{\srg/ART-XC discovery of pulsations from RX J0535.0-6700: another X-ray pulsar in the LMC}

   \author{I.~A.~Mereminskiy\address{1}\email{i.a.mereminskiy@gmail.com},
           A.~S.~Gorban\address{1,2}, Yu.~S.~Klein\address{2,1},
           E.~A.~Ushakova\address{2,1}, A.~N.~Semena\address{1},
           A.~A.~Lutovinov\address{1}, A.~Yu.~Tkachenko\address{1},
           S.~V.~Molkov\address{1}
     \addresstext{1}{Space Research Institute, Russian Academy of Sciences, Profsoyuznaya str., 84/32, 117997, Moscow, Russia}
     \addresstext{2}{National Research University 'Higher School of Economics', Pokrovsky bulvar, 11, 109028, Moscow, Russia}
    }

\shortauthor{Mereminskiy et al.}
\shorttitle{X-ray pulsations from RX J0535.0-6700} 

\submitted{01.04.2025}
\revised{07.04.2025}
\accepted{07.04.2025}

\begin{abstract}
    Using the {\it Mikhail Pavlinsky} ART-XC onboard the \srg\ observatory we have detected, for the first time, X-ray pulsations with a period of $\simeq106$ s from the poorly-studied high-mass X-ray binary \rx\, located in the Large Magellanic Cloud (LMC), thus proving that the accretor is a neutron star with strong magnetic field. Pulsations with similar period were also found in archival archival data from {\it Chandra} and XMM-{\it Newton} telescopes.
    Using photometry from {\it WISE} we shown that the source demonstrate significant variability in IR during the last twenty years, which could be caused by a secular evolution of the decretion disk.
    This discovery makes \rx\, another member of the large family of X-ray pulsars with Be-type companions in the LMC.
   
\end{abstract} 

\section*{Introduction}

The Large Magellanic Cloud (LMC) is a satellite dwarf galaxy of the Milky Way. Due to a powerful star formation outbursts in the recent past ($\approx 10^7$ years ago, see, e.g.  \citealt{2005A&A...431..597S, 2016MNRAS.459..528A}) it now harbours a large population of bright high-mass X-ray binaries (HMXBs). This population is very convenient to study with the modern X-ray telescopes due to a well-determined distance to LMC \citep[49.6 kpc, ][]{2019Natur.567..200P} and a moderate line-of-sight extinction \citep{2003MNRAS.339...87S}. 

However, the duty cycle of such systems is quite low, about 10\% or even less \citep{2018MNRAS.481.2779S,2018ApJ...868...47K} and major part of the time they reside in a low state with X-ray luminosity of $L_{\mathrm{X}}\lesssim10^{34}$ erg s$^{-1}$. Even for the most sensitive modern X-ray telescopes this makes them hard to study with observations of sensible exposures (dozens of ks). Therefore, in order to study these sources better, e.g. to measure neutron stars spin period, multiple observations are required, allowing one to catch new or known sources in an outburst.

\rx\, was discovered by the {\it ROSAT} observatory \citep{1982AdSpR...2d.241T} during the extensive LMC survey, composed from more than 200 observations performed from 1990 to 1994 \citep{1999A&A...344..521H}. The source was detected at the luminosity of $\approx3 \times 10^{35}$ \lum. 
\citet{1999A&A...344..521H} suggested that the optical counterpart of \rx\, is the bright blue star GRV 0535-6702, which was earlier classified as a Mira-type variable, given its flux variations with a period of 241 days \citep{1988MNRAS.232...53R}. Later, \citet{2002A&A...385..517N} performed optical spectroscopy of the candidate star and classified it as B0 Ve star, thus making \rx\ another Be-X-ray binary (BeXRB, \citealt{2012A&A...539A.114R}). In such systems a neutron star (NS) accretes matter from a decretion disc of the Be-star \citep[see, ][]{2009ApJ...707..870B}. Depending on orbital parameters an accretion could be either quasi-persistent \citep{2002ApJ...574..364P} or transient ones, with powerful outbursts occurring once per orbit or more rarely \citep[see, e.g., ][]{okazaki10}.

Using data from the {\it NTT} telescope \cite{2012A&A...539A.114R} measured the total equivalent width of the $H_{\alpha}$ line from the decretion disk in \rx\, to be -7.9\AA. Based on the $P_{orb} - EW(H_{\alpha})$ relation of BeXRBs  \citep{2011Ap&SS.332....1R} and assuming that the orbital period of \rx\, is $P_{orb} \simeq241$ days one can suggest that the decretion disk was far from its maximal size during the {\it NTT} observation in 2004. 

In 2024 \rx\, was serendipitously observed by the {\it Mikhail Pavlinsky} ART-XC telescope \citep{2021A&A...650A..42P} onboard the \srg\ observatory \citep{2021A&A...656A.132S} during long pointing observations of the another enigmatic LMC source 1A\,0538-66. These data allowed us to detect for the first time coherent pulsations with a period of $\simeq106$ s from the source. We identified this period as a spin period of the magnetic neutron star, thereby confirming classification of \rx\, as BeXRB. We have also looked through archival X-ray observations by the {\it Chandra} and XMM-{\it Newton} to trace the evolution of the spin-period and to study the source spectrum and its variability over the long timescales.

\section{Observations and data analysis}

The \rx\ happened to be located inside the ART-XC field of view during  monitoring observations of X-ray pulsar 1A\,0538-66 in the summer of 2024. Three long observations were carried out: first one in the beginning of June and two others in July (see  Tab.~\ref{tab:xobs} for details). In all three observations the source was detected at a significance level $>10\sigma$ in the standard $4-12$ keV band. 

\begin{table}[]
\caption{\label{tab:xobs} List of X-ray observations of \rx}
\centering
\begin{tabular}{lcc}
\hline\hline
ObsID & Start time, MJD & Exposure, ks\\
\hline
\multicolumn{3}{c}{ART-XC}\\
\hline
124101290010          &60464.86     & 173\\
124101290020          &60498.12     & 260\\
124101290030          &60508.20     & 86\\
\hline
\multicolumn{3}{c}{XMM-{\it Newton}}\\
\hline
0071740501            &52373.84     & 24\\ 
\hline
\multicolumn{3}{c}{{\it Chandra}}\\
\hline
27078                 &59948.23     & 25\\ 
27077                 &60147.66     & 28\\  
26555                 &60201.30     & 31\\  
28907                 &60536.70     & 18\\  
\hline
\end{tabular}

\end{table}

The ART-XC data was processed with the standard pipeline {\sc artproducts v0.9} using calibration data version {\sc CALDB} 20230228. For a temporal analysis we applied barycentric corrections to photon arrival times and extracted event lists and light curves for each observation from circular aperture with 1\arcmin radius, centred on the source position. The spectra were extracted from the smaller region with radius of $R=45\arcsec$. The background spectrum was obtained from earlier observations of a blank fields, and was normalized by the count rate at energies above 60 keV, where the effective area of the mirror system is negligible. Such approach allows to accumulate background spectra with better statistics and diminish the differences in response of individual detector pixels. The data from all seven telescope modules were combined for spectral and temporal analysis.

We also decided to look through the available archival data on the source, in order to trace its long-term behaviour. \rx\ was observed by the XMM-{\it Newton} observatory \citep{2001A&A...365L...1J} in 2002. 
Unfortunately, due to the observation mode, the source was observed only by the MOS1 and MOS2 cameras. It was located in outer chips,  which operated in the standard imaging mode, thus limiting temporal resolution to the frame integration time of 2.6 s. After standard data reprocessing using XMMSAS v20, we extracted event lists and spectra from the circular apertures of $R=22.5\arcsec$ ($R=20\arcsec$) for MOS1 (MOS2). The background spectra were extracted from the empty areas of the field on the same camera chips.

This part of LMC also was extensively covered with {\it Chandra} \citep{weisskopf2000} observations in 2023-2024. We selected four observations in which \rx\, was inside the telescope field of view and reprocessed data using \textsc{CIAO 4.17} \citep{2006SPIE.6270E..1VF}. Then we examined sky images in the {\it Chandra} broad band (0.5-7 keV). As it turned out, the source was securely detected in all of them, although it was the brightest in the last one, which was performed in 2024, nearly a month after the latest ART-XC observation. In this  observation {\it Chandra} accumulated about 1600 photons from the source, enabling to perform detailed spectral and temporal analysis. For each observation we extracted spectrum from a circular aperture with $R=6\arcsec$ centered on the source position; the background spectrum was obtained from empty parts of the field on the same CCD. For the last observation we have also extracted source lightcurve from the same circular aperture. Unfortunately, similar to XMM-{\it Newton}, the  temporal resolution for these data is also limited by the frame integration time, which is 3.1s. 

For spectral modelling we used the XSPEC v12.12.1 package \citep{1999ascl.soft10005A}. Given the low counting statistics we rebinned all spectra in order to have at least 5 counts per bin and applied W-statistics \citep{1979ApJ...230..274W} for fitting. Uncertainties on model parameters were derived by means of long Markov chains. To estimate a quality of best-fit models we employed the Cram\'er-von Mises statistics; for each model 1000 simulated datasets were produced, then for each dataset test statistic was calculated. A fraction of datasets with the lower (i.e. better) test statistic than for real data shows how well the model corresponds to the observed spectrum. For valid models this ratio should be about $50\%$. For all models, presented below, this ratio is below 60\%, indicating that this models are acceptable.

\section{Timing analysis}

\subsection{Pulsations}

\begin{figure}
\centering
   \includegraphics[width=1.\columnwidth]{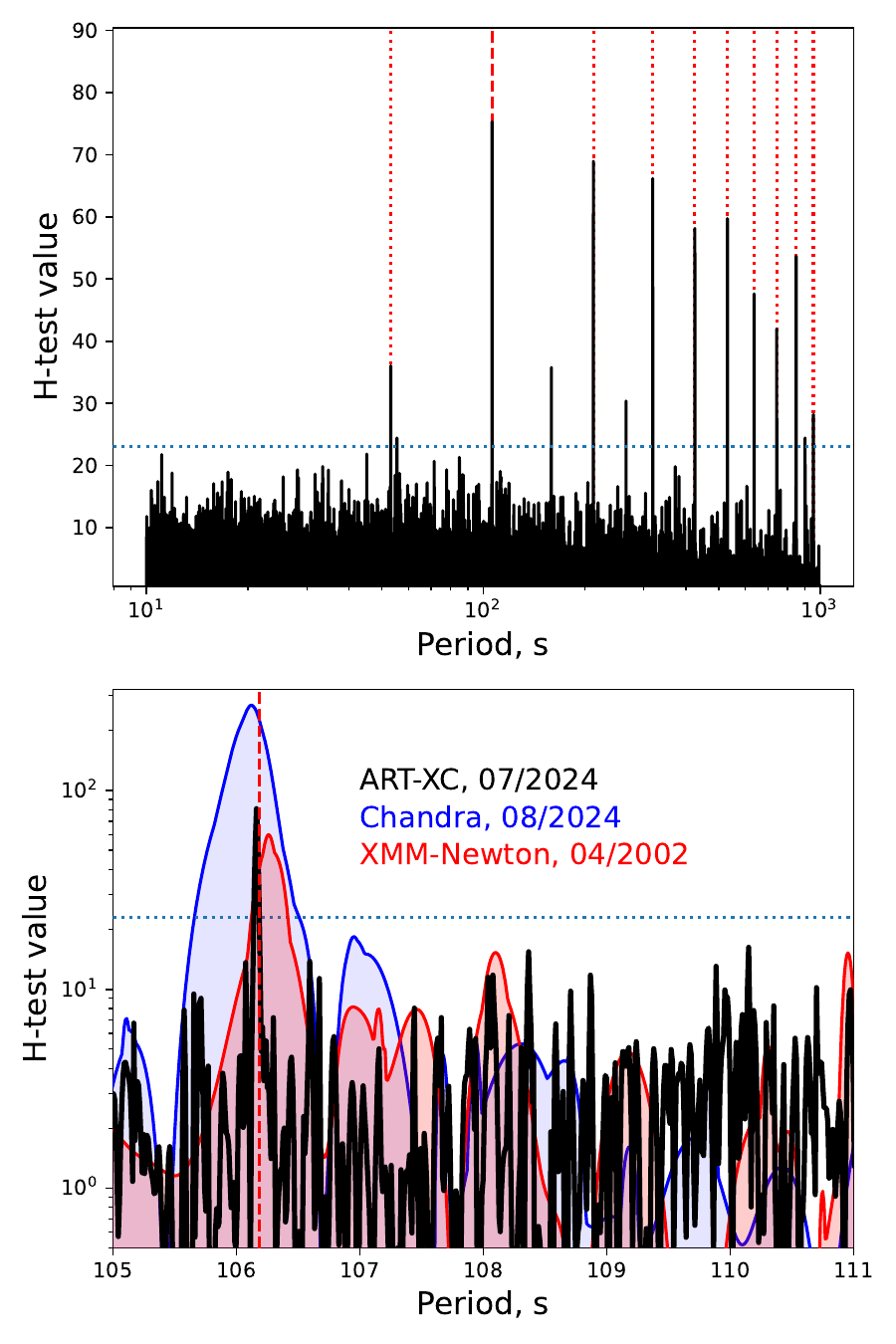}
     \caption{{\it Upper panel}: H-test periodogram of the first ART-XC observation (4-10 keV). Vertical red dashed line shows the fundamental period of the NS rotation, dotted lines shows harmonics and subharmonic. 
     {\it Lower panel}: part of the periodogram around the fundamental period from the second ART-XC observation (4-10 keV, black), XMM-{\it Newton} (0.2-10 keV, red) and {\it Chandra} (0.5-10 keV, blue) observations.}
     \label{fig:htest}
\end{figure}

Upon the identification of the observed source with the BeXRB candidate it became clear that ART-XC data could be used to search for the coherent pulsations from the source. From the event lists in the 4-10 keV band we calculated H-test \citep{hart1985choice} periodograms, utilising up to 20 harmonics \citep{1983A&A...128..245B}. The periodogram constructed on logarithmic grid of $4\times10^{4}$ test periods from 10 to 1000 s is shown in Fig.~\ref{fig:htest}. A horizontal line at $H=23$ corresponds to the detection of pulsations at a significance level of $>0.9999$. The peak at $P_{s}\approx106$ s clearly stands out, along with its harmonics  ($P = 0.5 P_{s}, 1.5 P_{s}, 2 P_{s}, 2.5 P_{s}, 3 P_{s}$ etc.). We identified this period as the spin period of neutron star in \rx.  

In a similar way we have produced periodograms for XMM-{\it Newton} and {\it Chandra} data. For each photon we generated a mock-up arrival time with a uniform random distribution inside the frame length. As it could be seen from the lower panel of Fig.~\ref{fig:htest} the pulsations with a nearly same period are detected in both datasets. Pulsations are also present in the second ART-XC observation, which was performed nearly a month after the first one. No periodic signal is detected at similar significance threshold in the third ART-XC observation.  

Although the periodogram is a reliable method to search for coherent signals in data, in order to study properties of such signals (e.g., exact measuring of the pulse period, pulse profiles) it is convenient to employ other methods. 

\subsection{Period determination and pulse profile}

In order to measure the spin period during each observation we constructed the 4-10 keV ART-XC lightcurves with the resolution of 1 s and used the epoch folding method \citep{leahy83}. Given the low count statistic (an observed ART-XC count rate is about 400 counts per day) to estimate uncertainty of the period measurement we used the technique, similar to one used by \cite{2022A&A...661A..33M}:  assuming that both background and mean over period source count rates were constant through the observation and using a piece-wise constant representation of the observed pulse profile we produced 1000 simulated lightcurves, that matches in the length and coverage with the real one. Then, for each simulated lightcurve we found the best period using the same epoch folding method. Boundaries of a confidence interval for the actual measurement was then calculated as 16\% and 84\% quantiles of sample of periods from simulated lightcurves. Applying this technique to the ART-XC data we obtained period measurements of $P = 106.180_{-0.005}^{+0.003}$ s (68\% confidence interval) and $P=106.160\pm 0.004$ s for first two observations, correspondingly. 

\begin{figure}
\centering
   \includegraphics[width=1.\columnwidth]{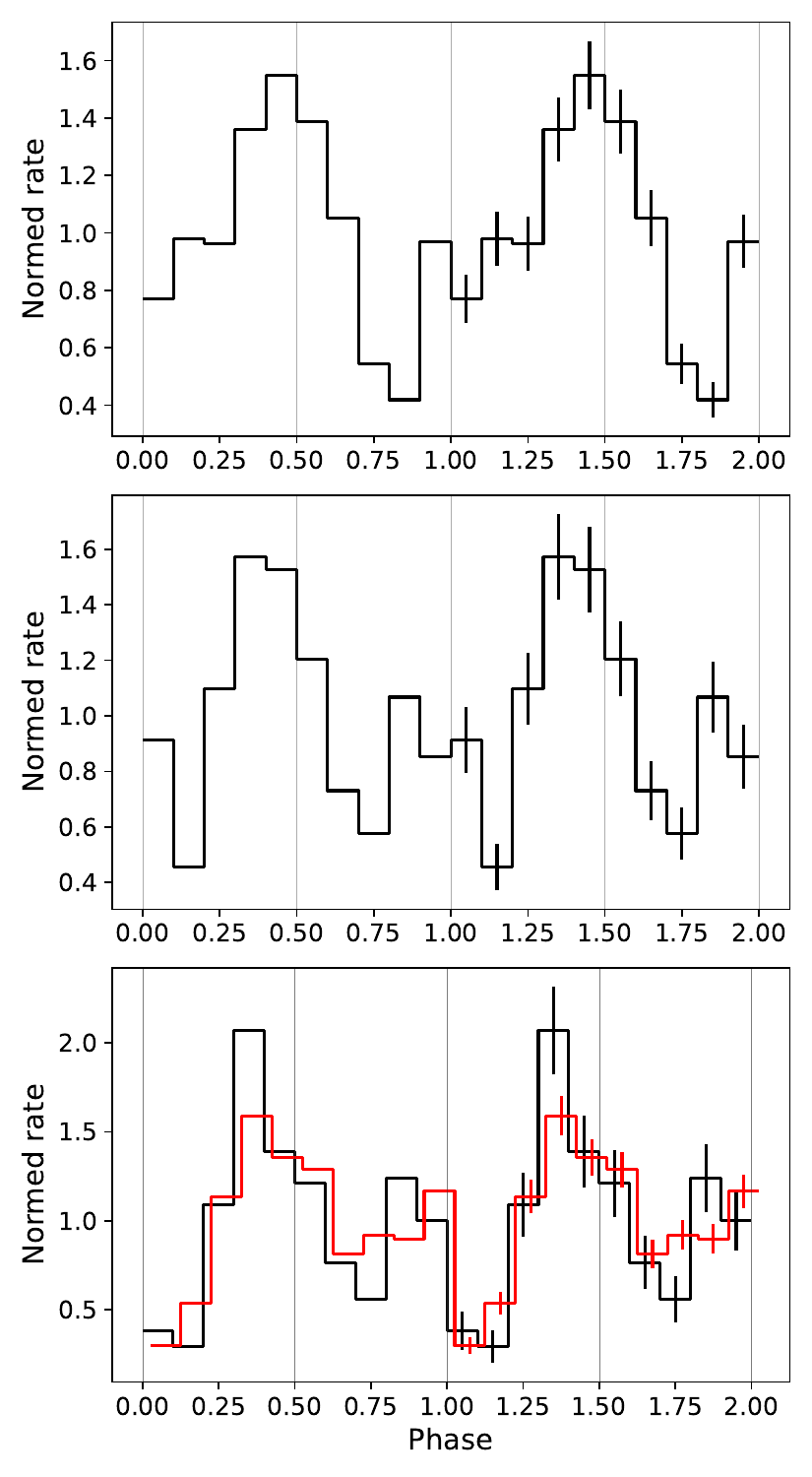}
     \caption{Pulse profile observed during different epochs. {\it Upper panel:} ART-XC. 4-10 keV, {\it middle:} XMM-{\it Newton}, 0.5-10 keV, {\it lower:} {\it Chandra} 0.5-2 keV (black) and 2-10 keV (red).}
     \label{fig:pprof}
\end{figure}

The pulse profile measured in the second ART-XC observation is shown in Fig.~\ref{fig:pprof} (upper panel). Formally, it has a simple one-peaked shape, but fine details could be lost due to insufficient statistics.  

For {\it Chandra} and XMM-{\it Newton} observations we used same technique, taking into account the corresponding frame accumulation times (3.1 and 2.6 s, correspondingly). For each event list we have produced 1000 realisations, assigning random arrival times inside the frame for each photon. Then we found best-fit period for each realisation.  Using the median and corresponding quantiles of sample of periods for each dataset, we derived $P=106.26^{+0.02}_{-0.02}$ s for the XMM-{\it Newton} observation in 2002 and $P=106.12\pm0.01$ s for the last {\it Chandra} observation in 2024. Corresponding pulse profiles from these datasets are shown in Fig.~\ref{fig:pprof} (middle and bottom panels). 

It is interesting to note that the pulsed fraction measured for all profiles is quite high: $\approx60\%$ for ART-XC and XMM-{\it Newton} and even higher $\approx70\%$ for {\it Chandra} data in narrow (0.5-2 and 2-10 keV) bands. 

\section{Spectral analysis}

In order to derive the source luminosity in our observations we investigated its X-ray spectra.

Given the relatively low  photons statistics we chose to use a simple phenomenological model consisting of the intrinsic X-ray continuum described by a cut-off powerlaw and two absorption components -- one due to the local absorption near the source and another one that corresponds to the line-of-sight absorption in the Milky Way: \texttt{tbabs*tbfeo*cutoffpl} in XSPEC terms. The thickness of the neutral hydrogen in the Galaxy in the source direction was fixed at the value of $N_{\mathrm{H,\,loc}}=10^{21}$ cm$^{-2}$ \citep{HI4PI}. Since the LMC metallicity significantly differs from the Solar one, following  \cite{ducci2019awakening} for the local absorption we used oxygen and iron abundances of 0.33 and 0.38, respectively.

\begin{figure}
\centering
   \includegraphics[width=1.\columnwidth]{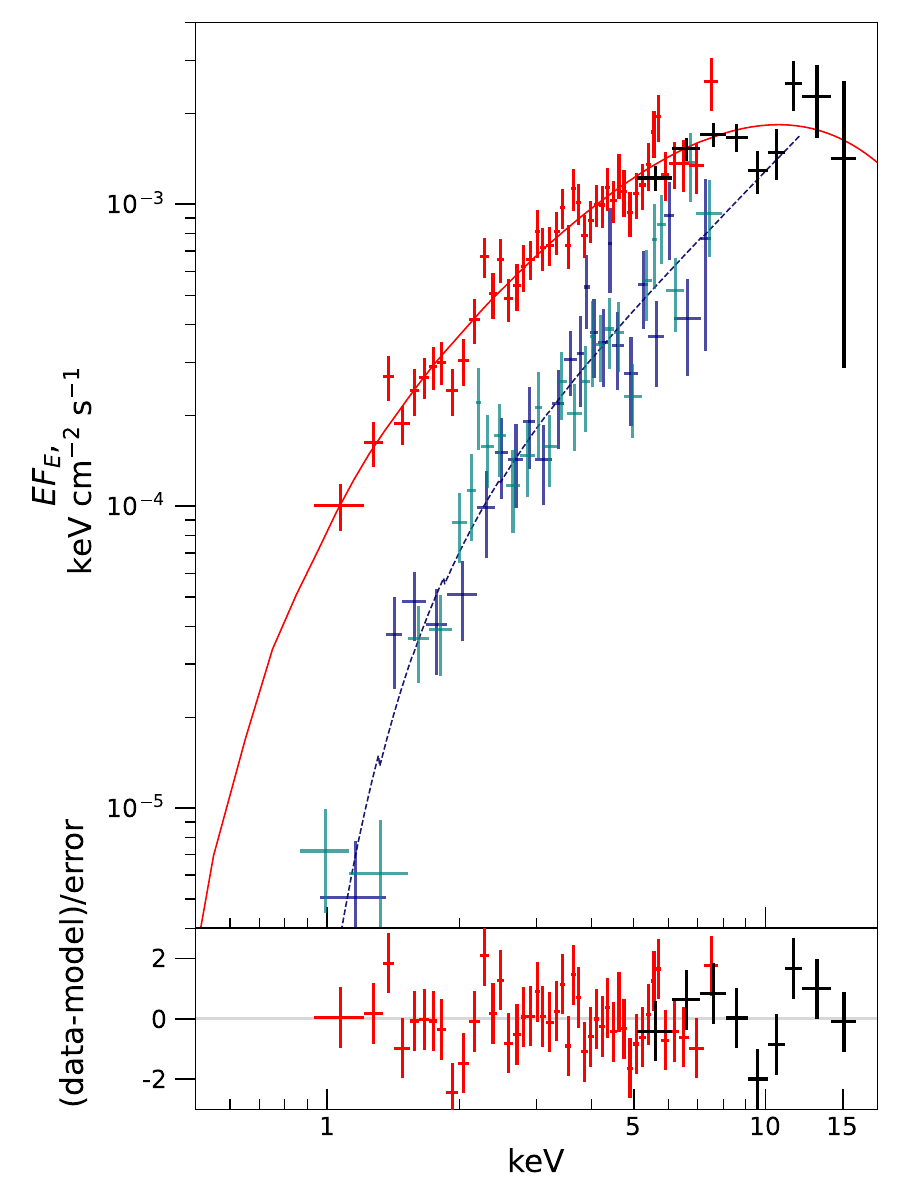}

       \caption{X-ray spectra obtained in 2024 by {\it Chandra} (red crosses) and ART-XC (black crosses, scaled to match {\it Chandra} data), shown along with the best-fit model (solid line) and corresponding residuals. Navy and cyan points shows the spectrum from XMM-{\it Newton} MOS1/MOS2, accumulated in 2002. }
     \label{fig:xspec}
\end{figure}

Despite the model simplicity, it is not trivial to robustly estimate its parameters from a narrow-band spectrum due to arising degeneracies. Therefore we decided to simultaneously fit spectra from first two ART-XC observations and {\it Chandra} spectrum, which was obtained nearly a month later. Two ART-XC spectra are similar in shape, and observed 4-12 keV count rates differs by less then $10\%$. However, during the {\it Chandra} observation the source was significantly brighter. To account for a differences in normalization we added multiplicative constant to the model. The resulting broadband X-ray spectrum (0.8-16 keV) is shown in Fig.~\ref{fig:xspec} by red and black crosses. Thanks to its extended coverage we were able to measure an intrinsic absorption of $N_{\mathrm{H,\,loc}} = 2.8^{+7.5}_{-2.1} \times 10^{21}$ cm$^{-2}$ (hereafter uncertainties on parameters corresponds to 90\% level), a power-law slope of $\Gamma = 0.23_{-0.17}^{+0.52}$ and a cut-off energy of $E_{\mathrm{cut}} = 6.0_{-1.2}^{+7.2}$ keV.

\begin{figure}
\centering
   \includegraphics[width=1.\columnwidth]{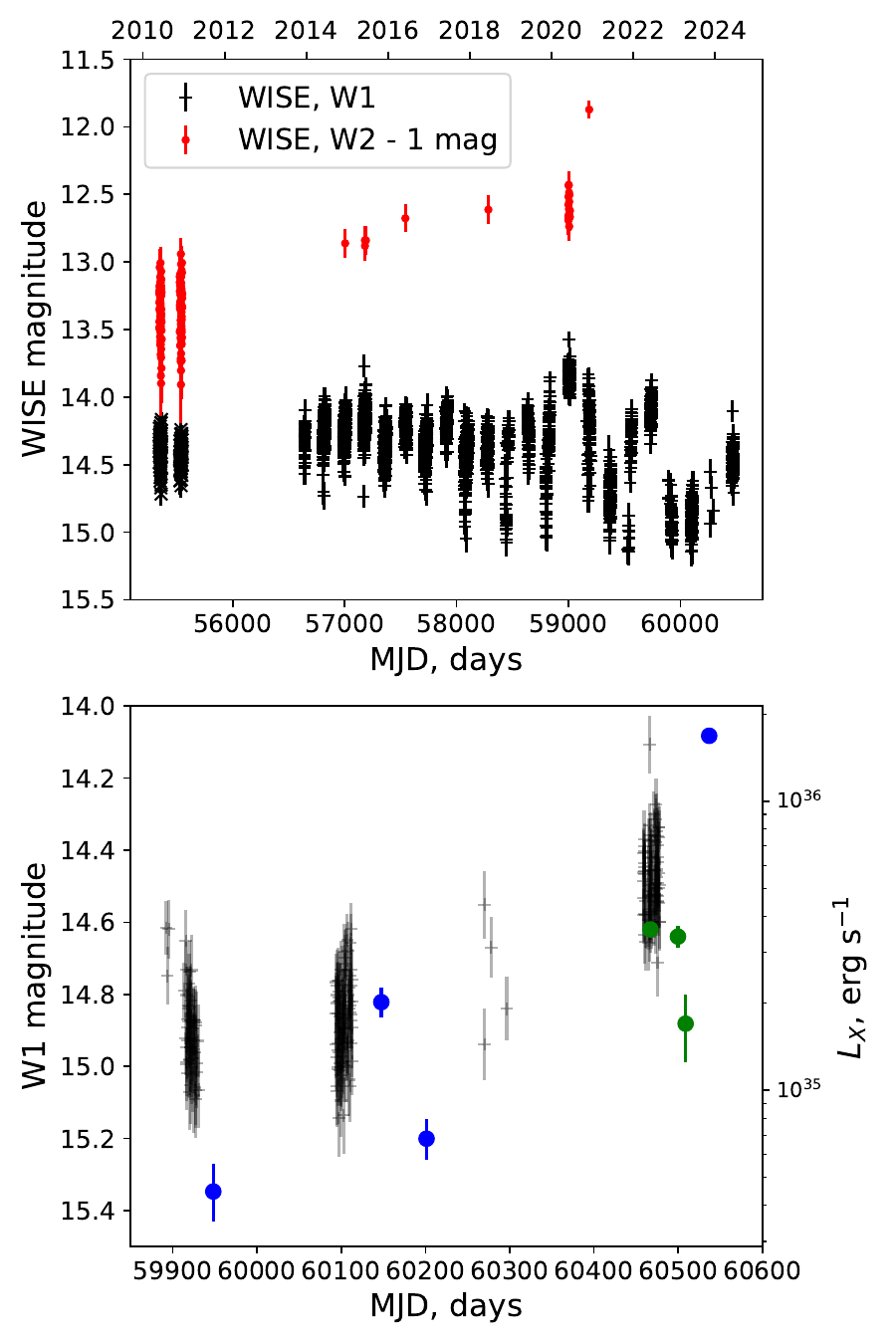}

     \caption{{\it Upper panel:} 2010-2024 IR lightcurve of \rx\, in WISE W1 (shown in black) and W2 (red, shifted by 1 mag) filters.\\
     {\it Lower panel}: zoom-in on 2023-2024 WISE W1 data. X-ray luminosities of \rx\, from ART-XC and {\it Chandra} observations shown with green and blue circles, correspondingly.}
     \label{fig:wiselc}
\end{figure}

We used these parameters to estimate a bolometric 0.1-30 keV luminosity for all ART-XC and {\it Chandra} observations in 2023-2024. As it is shown in Fig.~\ref{fig:wiselc}, over two years the source have brightened from $4\times10^{34}$\lum\, up to $2\times10^{36}$ \lum, although the growth was not monotonic, with an apparent variability on timescales of weeks. 

Interestingly, while the archival spectrum from the 2002 XMM-{\it Newton} observation could be described with the similar spectral model with the intrinsic absorption fixed at $N_{\mathrm{H,\,loc}} = 2.8\times 10^{21}$ cm$^{-2}$ it leads to unphysical continuum parameters, such as the extremely hard power-law slope of $\Gamma\lesssim -1.5$ and low cut-off energy $E_{cut}\approx3$ keV. The spectrum could also be described with a single blackbody model with the temperature of $kT\approx2$ keV and characteristic size of the emission region of $R\approx500$ m. In this case the total luminosity turns out to be around $3\times10^{35}$ \lum. Note that such hotspots are often observed in X-ray pulsars, albeit at much higher luminosities -- typically above $10^{37}$ \lum \citep{2022arXiv220414185M}.

However, by allowing a local absorption to vary freely we obtained a satisfactory fit with a moderate power-law index $\Gamma = 0.49_{-0.25}^{+0.47}$ and much larger $N_{\mathrm{H,\,loc}} = 23.5^{+17.0}_{-7.1} \times 10^{21}$ cm$^{-2}$. The spectral coverage is not enough to constrain the cut-off energy. This model, along with the unfolded MOS1/MOS2 data, shown in Fig.\ref{fig:xspec} by a navy line and navy and cyan crosses. It is impossible to get a meaningful estimate of the bolometric luminosity for such a hard spectrum. But using the soft X-ray luminosity of $L_{0.1-10\,keV}\approx4\times10^{35}$ \lum and assuming that a bolometric correction  factor is moderate ($L_{0.1-30\,keV}/L_{0.1-10\,keV}\approx1.7$ for a broadband ART-XC/{\it Chandra} spectrum) one can guess that the total bolometric luminosity was about $10^{36}$ \lum at the moment of the XMM-{\it Newton} observation in 2002.

It is interesting to note that XMM-{\it Newton} observed the field of \rx\, several times on May - June, 2018, but the source was not detected. We queried the  XMM-{\it Newton} Science Archive upper limit server \citep{2022MNRAS.511.4265R} and found that a typical 2$\sigma$ upper limit on the source flux is $\approx 3\times10^{-14}$ \flx\ in the 0.2-12 keV energy range. This value corresponds to the intrinsic luminosity of $\approx10^{34}$ erg s$^{-1}$, assuming that the spectral shape was similar to the one observed in 2024.

\section{Discussion}

The long-term variability of the X-ray luminosity in Be-systems, such as observed in \rx, could be related to the secular evolution of the decretion disk \citep{2010ApJ...709.1306W}. We examined the publicly available data from the IR space telescope {\it WISE} \citep{2010AJ....140.1868W, 2014ApJ...792...30M} to search for the presence of a characteristic ''brighter-redder'' variability, which is related to variations of the decretion disk size. Light curves in filters W1 and W2 are shown in Fig.~\ref{fig:wiselc}. Unfortunately, amount of the W2 data is not enough for definitive conclusions about the state of the decretion disk in 2023-2024. But it could be seen, that the brightness in the W2 filter has grown from 14.5 to 13 magnitude since 2010; additionally the W1 magnitude was also decreased by $\approx1$ mag. Also, there is an apparent change in the variability pattern around 2019. Further spectroscopic observations are necessary to probe the decretion disk state in this system. 

If the optical period of 241 days \citep{1988MNRAS.232...53R} is indeed the orbital one, then a detected spin-period of $P\simeq106$ s puts \rx\, on a region of the Corbet diagram  \citep{1986MNRAS.220.1047C} occupied by other BeXRBs \citep[see, updated versions of the diagram in][]{2010arXiv1004.0293G,2019NewAR..8601546K}.   

The stability of the observed NS spin-period over the las twenty years implies that it is close to the equilibrium one. Assuming that the mean mass accretion rate in \rx\, is $\langle \dot M\rangle \approx2\times10^{-11}\,M_{\odot}$ yr$^{-1}$, which corresponds to mean luminosity of  $\langle L_{X}\rangle \approx10^{35}$ erg s$^{-1}$, one could estimate the NS magnetic field strength using Eq. 9 from \citealt{1997ApJS..113..367B}: $B\approx5\times10^{12}$ G. This is not enough for the ''propeller'' effect \citep{1975A&A....39..185I} to set on, allowing the accretion to take place through a cold disk \citep{2017A&A...608A..17T}. This could explain a long lasting episode with low luminosity ($L_X\approx3\times10^{35}$ erg s$^{-1}$) observed by ART-XC, which is not typical for a BeXRB variability.

\section{Conclusion}

We have discovered coherent X-ray pulsations with a period of 106.2 s from \rx\, -- a BeXRB located in the LMC. Pulsations present in the ART-XC data, in which they were revealed initially, as well as in XMM-{\it Newton} and {\it Chandra} data, obtained during different epochs. The pulse period remains relatively stable in all observations, spanning more than 20 years, and the pulsed fraction is quite high also in all observations, above 50\% in all energy bands.

The source X-ray spectrum is typical for accreting X-ray pulsars and can be described by an absorbed power law with an exponential cutoff. It seems that the local absorption has changed dramatically between 2002 and 2024, although it should be mentioned that the accurate measurement of $N_{\mathrm{H,\,loc}}$ in the standard X-ray band (0.5-10 keV) is complicated. The broadband observations are required to better characterize spectral parameters.

The optical companion of \rx\, B0Ve star exhibits a strong variability in the IR bands, which could indicate the secular variations of the decretion disk. 

Finally we can conclude that all obtained data confirms that \rx\ is the 28th member of a large family of BeXRBs with neutron stars in the LMC \citep{haberl2023lmc}.

\acknowledgements

This work is based on observations with the Mikhail Pavlinsky ART-XC telescope, hard X-ray instrument on board the SRG observatory. The SRG observatory was created by Roskosmos in the interests of the Russian Academy of Sciences represented by its Space Research Institute (IKI)
in the framework of the Russian Federal Space Program, with the participation of
Germany. The ART-XC team thanks Lavochkin Association (NPOL) with partners for the creation and operation of the SRG spacecraft (Navigator)

This research has made use of data obtained from the Chandra Data Archive provided by the Chandra X-ray Center (CXC).

Research is partially based on observations obtained with XMM-Newton, an ESA science mission with instruments and contributions directly funded by
ESA Member States and NASA

This work was supported by the Minobrnauki RF grant 075-15-2024-647.


\label{lastpage}


\bibliographystyle{mnras} 
\bibliography{reflist_rx} 

\end{document}